\documentstyle[aps,prb,epsbox]{revtex}

\draft
\begin{document}


\title{Renormalized Bosonic Interaction of Excitons
}
\author{Jun-ichi Inoue\cite{email1}, 
Tobias Brandes\cite{newaddress}, 
and Akira Shimizu\cite{email2}}
\address{Department of Basic Science, University of Tokyo,
3--8--1 Komaba, Meguro-ku, Tokyo 153--8902, Japan
\\
and Core Research for Evolutional
Science and Technology, JST}
\date{\today}
\maketitle

\begin{abstract}
An effective bosonic 
Hamiltonian of $1s$ excitons with ``spin''
 degrees of freedom in 
two dimension is obtained through a projection procedure,
starting from a conventional electron-hole Hamiltonian ${\cal H}_{eh}$. 
We first demonstrate that
a straightforward transformation of ${\cal H}_{eh}$ 
into a Hamiltonian of bosonic excitons does not 
give the two-body interaction between an ``up-spin'' exciton 
and a ``down-spin'' exciton, which are created by
the left- and right-circularly polarized light beams, respectively.
We then show that this interaction is generated through 
a projection procedure onto the subspace spanned by $1s$ excitons,
as a renormalization effect coming from higher exciton states.
The projection also renormalizes the interaction 
between $1s$ excitons with the same spins by a large amount.
These renormalization effects are
crucial for the polarization
 dependence of the optical responses from semiconductors.
The present theory gives the microscopic foundation of the phenomenology 
that was successfully applied to 
the analysis of four-wave mixing experiments in GaAs 
quantum wells strongly coupled
 to the radiation field in a high-Q micro cavity.
\end{abstract}

\pacs{PACS numbers: 71.35.-y, 71.10.-w, 78.66.-w}


\section{Introduction}
Various theoretical methods have been developed to study 
optical properties of
semiconductors \cite{Haug,Lindberg,Hanamura,Axt,Ivanov}.  
These methods can be divided into two groups; 
a ``fermionic method'' and a ``bosonic method''.  
%
The fermionic method \cite{Haug,Lindberg} is 
formulated on the Hilbert space of two fermionic
species, i.e., photo-excited electrons and
holes in semiconductors.  In this method, one solves the
coupled equations of motions, the semiconductor Bloch
equations (SBE) \cite{Lindberg}, for the particle densities of electrons
and of holes, and for the expectation value of the polarization of the
system.  
Since this method basically relies on 
the Hartree-Fock (HF) approximation for electrons and holes, 
it is suitable for
higher excitation density, where Coulombic screening effects guarantee that
exciton correlation effects become less crucial.
In order to extend this method to lower excitation density, 
where exciton correlations become important, 
the three- and four-particle correlations should be taken into 
account and the truncation scheme should be improved.

On the other hand, 
the bosonic method \cite{Hanamura,Axt,Ivanov} is based on excitons,
bound states of an electron and a hole.
Regarding this elementary excitation as a bosonic particle, 
one constructs an effective Hamiltonian of bosonic excitons, 
from which physical quantities, such as linear and non-linear response functions, can be calculated.
Here, the effective Hamiltonian should be constructed very carefully, 
as we will show in this paper.
The bosonic method is believed to be valid 
when the optical excitation is weak 
and when the photon energy is close to
the exciton energy (see section \ref{conditions}), 
because under these conditions main
contributions to the optical properties 
should come from the excitons created in the system.  
With increasing the photo-excitation intensity, 
the Coulombic force becomes weaker by increased screening and/or increased Fermi energy, and the contribution from free carriers becomes more
important. Hence, the bosonic method is 
not valid at high excitation intensity.  


A remarkable 
feature of the light field as a probing tool of materials is
that it has the polarization degrees of freedom.  
Recent experimental studies of semiconductor optics make
the best use of this fact to reveal more detailed properties of
excited states.  
The polarization degrees of freedom of photons 
induce ``spin'' degrees of freedom
of excitations (see below).  
Experiments, 
including the four-wave mixing experiment in the time-domain, 
have revealed the crucial roles of 
the interaction between an exciton (or an $eh$ pair) 
created by the left-circularly
polarized light and the one created by the right-circularly
polarized light\cite{Bott,Schafer}.
%
However, 
most of the existing theories 
could not treat the polarization dependence
correctly.  
For example, since the SBE (in its original form) 
were discussed within HF theory \cite{Lindberg}, 
the excitations with different polarization degrees of freedom
(e.g., left and right) are completely decoupled.  
Hence it is impossible to explain the polarization dependence of the
optical response.  
It is the case with bosonic theories.  
In most cases, the polarization degrees of freedom were not included in
the theories for simplicity.  
Although it has been asserted that including such degrees
of freedom should be trivial \cite{Hanamura}, the excitons
with opposite spins are completely decoupled.  
This is  because
they essentially end up with HF of $1s$ excitons.  
This means that to discuss polarization dependence in bosonic theories
is non trivial.  

In this paper, we show that
it is not at all trivial to derive the interaction of excitons created by opposite polarized
light in a bosonic theory.  
One crucial result is that to go beyond the HF approximation of $1s$
excitons becomes of paramount importance.  
One of such investigations is seen in Ref.\,\cite{Fernandez}, where
the modification of exciton binding energy is discussed.

Up to now, a strong objection against the bosonic
method was that excitons are not bosonic particles.
Moreover, until recently there were almost no experimental evidence
which  verifies an effective
bosonic theory both qualitatively and quantitatively.  
However, important experimental evidence for the validity of the
boson picture has been reported recently in a two-dimensional system
\cite{Kuwata-Gonokami,Shirane}.
The experiment, discussed in detail below, is a non-linear version of normal mode
coupling in a high-Q micro cavity.  
This experiment was stimulated by the fact that
optical responses from semiconductors
that are strongly coupled to the photon field have received much
attention in recent years.  
The strong coupling is obtained by confining optically active
regions in a high-Q micro cavity, which is made
possible by the development of nano-structure technology.
For linear optical responses,
the most noticeable phenomenon is
a large Rabi-splitting, which has been observed in
both inorganic \cite{Weisbuch} and organic\cite{Lidzey}
semiconductors (insulators). 

Kuwata-Gonokami {\em et al.} have performed four-wave mixing
in a GaAs quantum well (QW) that is strongly coupled to the 
radiation field in a
high-Q micro cavity\cite{Kuwata-Gonokami,Shirane}.
They investigated several polarization
configurations under
the condition that the excitation density is very low, 
and measured the polariton-polariton scattering signals.
The experimental results were successfully
reproduced by a phenomenological model, which is called 
the weakly interacting boson model (WIBM)\cite{Kuwata-Gonokami,Shirane,Suzuura}.
In the WIBM, excitons are treated as
interacting bosons.  
The good agreement with the experimental
results demonstrated that the 
bosonic picture is reliable in the lower-excitation regime.
The Hamiltonian of the WIBM is
\begin{eqnarray}
 {\cal H}_{\rm WIBM}&=&\sum_{{\bf k},\sigma}
\left[\omega_{c}a^{\dagger}_{{\bf k}\,\sigma}a^{\phantom{\dagger}}_{{\bf k}\,\sigma}
     +\omega_{e}b^{\dagger}_{{\bf k}\,\sigma}b^{\phantom{\dagger}}_{{\bf k}\,\sigma}
     +g(a^{\dagger}_{{\bf k}\,\sigma}b^{\phantom{\dagger}}_{{\bf k}\,\sigma}
     +b^{\dagger}_{{\bf k}\,\sigma}a_{{\bf k}\,\sigma})\right] 
 \nonumber\\
&&+W\sum_{{\bf k}_{1},{\bf k}_{2},{\bf k}_{3}}
b^{\dagger}_{{\bf k}_{1}\,+}b^{\phantom{\dagger}}_{{\bf k}_{2}\,+}
 b^{\dagger}_{{\bf k}_{3}\,-}b^{\phantom{\dagger}}_{{\bf k}_{1}-{\bf k}_{2}+{\bf k}_{3}\,-}
\nonumber\\
&&+R\sum_{{\bf k}_{1},{\bf k}_{2},{\bf k}_{3},\sigma}
b^{\dagger}_{{\bf k}_{1}\,\sigma}
b^{\dagger}_{{\bf k}_{2}\,\sigma}
b^{\phantom{\dagger}}_{{\bf k}_{3}\,\sigma}
b^{\phantom{\dagger}}_{{\bf k}_{1}-{\bf k}_{2}+{\bf k}_{3}\,\sigma}
\nonumber\\
&&-g\nu\sum_{{\bf k}_{1},{\bf k}_{2},{\bf k}_{3},\sigma}
\left(
b^{\dagger}_{{\bf k}_{1}\,\sigma}
b^{\phantom{\dagger}}_{{\bf k}_{2}\,\sigma}
a^{\dagger}_{{\bf k}_{3}\,\sigma}
b^{\phantom{\dagger}}_{{\bf k}_{1}-{\bf k}_{2}+{\bf k}_{3}\,\sigma}
+
b^{\dagger}_{{\bf k}_{1}\,\sigma}
a^{\phantom{\dagger}}_{{\bf k}_{2}\,\sigma}
b^{\dagger}_{{\bf k}_{3}\,\sigma}
b^{\phantom{\dagger}}_{{\bf k}_{1}-{\bf k}_{2}+{\bf k}_{3}\,\sigma} 
\right),
\label{wibm}
\end{eqnarray}
where photons and excitons are described by boson operators,
$a_{{\bf k}\,\sigma}$ and $b_{{\bf k}\,\sigma}$, respectively, 
with the spin index
$\sigma=\pm$.
This Hamiltonian has three parameters: $W$ is the interaction strength
of the excitons with opposite spin, $R$ is the interaction strength
of the excitons with the same spin, and $\nu$ is the filling factor \cite{Schmitt-Rink2}.  
The ratio of these three parameters was measured as
$R:W:g\nu=-3.0:0.2:1.0$.  
Once these three parameters are fixed, the experimental
results for all polarization configurations are fitted very well.  
This indicates
that a bosonic method is quite reliable in the lower excitation density.
Nevertheless, existing bosonic theories \cite{Hanamura,Axt,Ivanov} cannot
give a {\em microscopic} foundation of the WIBM.  

The purpose of our paper is to derive an effective boson Hamiltonian of
$1s$ excitons with ``spin'' degrees of freedom under the three
conditions cited in Sec.\,\ref{conditions}.  
Through this derivation, a microscopic foundation of the WIBM, especially the interaction terms of
excitons with the scattering strength $W$ and $R$ in the WIBM, is
obtained.  
We show that effects of 
exciton states higher than $1s$ are crucial 
when deriving the effective Hamiltonian of $1s$ excitons \cite{Inoue}:
these effects yield the two-body
interaction term corresponding to $W$ in Eq.\,(\ref{wibm}),
and largely modify the strength of 
the interaction term corresponding to $R$ in Eq.\,(\ref{wibm}).
The higher exciton states are taken into account through 
a projection procedure, and, 
as shown in Sec.\,\ref{necessity},
it is crucial to project the whole exciton space onto the $1s$ excitons
subspace and to re-construct of the interaction of $1s$ excitons.

In a more general sense, the bosonic description of fermionic systems in two-dimension is one of
the most attractive fields in condensed matter theory \cite{Stone,Kopietz}.
If the bosonic method is useful in two-dimensional semiconductors, it
would be one of the examples of the success of a two-dimensional
``bosonization''.

The organization of this paper is as follows:
In Sec.\,II, after explaining the necessity of the projection, exciton
bosonic operators are introduced and exciton spins
are defined.  
In Sec.\,III, the exciton interaction is discussed without the
projection, which corresponds to the two-dimensional case of the
existing bosonic theory \cite{Hanamura}.  
It is shown that the exciton interaction corresponding to the $W$ term in
WIBM is {\em not} obtained in this approximation.  
In Sec.\,IV, which is the main part of this paper, the projection
procedure is introduced \cite{Inoue}, resulting in appearance of the $W$ term and modifications of the
interaction strength of excitons with the same spins.  
Discussions and remarks are presented in Sec.\,V.  
In Sec.\,VI, the results obtained in this paper are summarized.  
Throughout this paper, the units  $\hbar=1, e=1, \epsilon_{0}=1$ are used.

\section{Model and strategy}

\subsection{Conditions for the validity of the effective theory}
\label{conditions}

The purpose of the present paper is 
to derive an {\em effective} Hamiltonian of $1s$ excitons
in a QW.
Here, the ``effective Hamiltonian'' means that 
it (approximately) 
describes the optical responses of the QW correctly, 
although it is a function of $1s$ exciton operators only.

Since any effective theory is valid only in some specific
physical situations,
we first clarify the physical situations or conditions  
under which we construct the effective theory.
In compensation for this limitation, the effective theory is quite 
useful; it gives deep insights by describing the physics simply.
In contrast, such insights are hardly obtainable from 
straightforward calculations using the electron-hole Hamiltonian.

We construct an effective Hamiltonian that describes 
optical responses of semiconductor QWs 
under the following conditions: 
(i) The excitation is weak (weak excitation regime)
so that 
the mean distance $l_{ex}$ of photo-created 
(virtual and/or real) excitons is much
larger than the Bohr radius $a_{0}$ of the $1s$ exciton;
\begin{equation}
l_{ex} \gg a_{0},
\label{weak}\end{equation}
and 
(ii) all the photon energies (pump, probe, and signal) 
$\hbar \omega_i$ are close to
the energy $E_{1s}$ of the $1s$ exciton; 
\begin{equation}
|E_{1s}-\hbar \omega_i|
\ll
|E_{2p}-\hbar \omega_i|
\quad \mbox{($i=$ pump, probe, signal)}
\label{close}\end{equation}
where $E_{2p}$ is the energy of $2p$ excitons,
and
(iii)
the line width $\Gamma_{1s}$ of $1s$ exciton is smaller than the 
detuning energies;
\begin{equation}
\Gamma_{1s}<|E_{1s}-\hbar \omega_i|.  
\quad \mbox{($i=$ pump, probe, signal)}
\label{width}\end{equation}
Physical meaning of these conditions will be explained
in section \ref{sec_validity}.

Under these conditions, nonlinear optical signals would not be strong in general.
One must therefore devise experimental methods for 
detecting the signals with a high sensitivity.
For this purpose, a genius method was proposed by Kuwata-Gonokami 
{\em et al.} \cite{Kuwata-Gonokami}, in which an optical cavity with a high Q value
is utilized.
This point will be discussed later in Sec.\,\ref{sec_validity}.

\subsection{Necessity of projection and renormalization}
\label{necessity}

The Hamiltonian of an $eh$ system is defined on 
the $eh$ Hilbert space ${\sf H}_{eh}$ that is an fermionic 
Hilbert space spanned by $e$ and $h$ states.
As long as photo excitations are concerned, all excited states are charge neutral. We can thus limit ourselves 
in the charge neutral sector of ${\sf H}_{eh}$.
The effective Hamiltonian of excitons, which is 
defined on 
a bosonic Hilbert space spanned by the exciton states, 
should describe the dynamics of the $eh$ system 
in this charge neutral sector of ${\sf H}_{eh}$.
Note that there are two (or more) choices 
for the bosonic Hilbert space: 
one is the whole exciton space ${\sf H}_{ex}$ that is spanned by all
exciton states, whereas 
the other is its subspace  ${\sf H}_{1s}$ that is spanned by 
the $1s$ states only.
The effective Hamiltonian depends on the choice of 
the bosonic Hilbert space.

We will first consider in Sec.\,\ref{before}
the effective Hamiltonian defined on ${\sf H}_{ex}$.
Its interaction part 
takes the following form:
\begin{equation}
{\cal H}^{full-int}
=
\frac{1}{2\Omega}\sum_{\{S\}}\sum_{\{\nu\}}\sum_{{\bf k},{\bf k}',{\bf q}}
{V}_{ex}({\bf q};\{\nu\};\{S\})
b^{\dagger}_{{\bf k}+{\bf q}\,\nu_{1}\,S_{1}}
b^{\dagger}_{{\bf k}'-{\bf q}\,\nu_{2}\,S_{2}}
b_{{\bf k}'\,\nu_{4}\,S_{4}}
b_{{\bf k}\,\nu_{3}\,S_{3}}
+ \cdots,
\label{generalH}
\end{equation}
where $ b_{{\bf k}\,\nu\, S}$ denotes 
the excitonic boson operators (defined later), 
and $\Omega$ is the area of the QW.
The first term denotes the two-body interactions between 
excitons of various states,
and $\cdots$ 
denotes three- and more- body interactions.
%
Since {\em exact} calculations, which take account of {\em all} 
terms of Eq.\,(\ref{generalH}) to infinite order, are impossible, 
{\em one has to make approximations}.
Because of conditions (i)-(iii) of Sec.\,\ref{conditions}, 
it is tempting to take only the two-body interactions of 
$1s$ excitons,
${\cal H}^{int}_{1s}$,
among many terms of Eq.\,(\ref{generalH});
\begin{equation}
{\cal H}^{int}_{1s}
=
\frac{1}{2\Omega}\sum_{\{S\}}\sum_{{\bf k},{\bf k}',{\bf q}}
{V}_{ex}({\bf q};\{\nu=1s\};\{S\})
b^{\dagger}_{{\bf k}+{\bf q}\,S_{1}}b^{\dagger}_{{\bf k}'-{\bf q}\,S_{2}}b_{{\bf k}'\,S_{4}}b_{{\bf k}\,S_{3}},
\label{generalHs}
\end{equation}
where $b_{{\bf k}\,S}$ denotes $b_{{\bf k} \, \nu \, S}$ with 
$\nu=1s$ \cite{short-range}.
Unfortunately, however, 
we will show later that the replacement 
 ${\cal H}^{full-int}$ $\to$ ${\cal H}^{int}_{1s}$ is 
a very poor approximation, which cannot explain 
the experimental results {\em even qualitatively}.
This originates from the complete neglect of 
effects of higher exciton states $\nu=2p, 3d,\ \cdots$,
which, however, play important roles as intermediate states.

To resolve this difficulty, we will then 
consider in Sec.\, \ref{projection}
the effective Hamiltonian defined on ${\sf H}_{1s}$.
It is obtained by the projection procedure, 
by which 
the dynamics in ${\sf H}_{ex}$ is projected onto 
the subspace ${\sf H}_{1s}$ that is spanned by $1s$ excitons only.
In general, a projection procedure generates dissipative terms in 
the projected dynamics in the subspace.
Under conditions (i)-(iii) of Sec.\,\ref{conditions}, however, 
we may neglect the dissipative terms.
Namely, the dynamics in ${\sf H}_{1s}$ can be described 
by an Hamiltonian dynamics, whose Hamiltonian 
(effective Hamiltonian) is a function of 
the boson operators $b_{{\bf k}\,S}$ for $1s$ excitons only.
Its interaction part, 
$\tilde{\cal H}^{full-int}_{1s}$,
consists of two- and more-body interactions among $1s$ excitons.
For example, the two-body interaction $\tilde{\cal H}^{int}_{1s}$
takes the following form;
\begin{equation}
\tilde{\cal H}^{int}_{1s}
=
\frac{1}{2\Omega}\sum_{\{S\}}\sum_{{\bf k},{\bf k}',{\bf q}}
\tilde{V}_{ex}({\bf q};\{S\})
b^{\dagger}_{{\bf k}+{\bf q}\,S_{1}}
b^{\dagger}_{{\bf k}'-{\bf q}\,S_{2}}
b_{{\bf k}'\,S_{4}}
b_{{\bf k}\,S_{3}}.
\label{projectHs}
\end{equation}
In this two-body interaction, 
effects of higher exciton states 
have been (partly) included as ``renormalization effects'', 
which have modified (renormalized) the forms and the strengths
of $\tilde{\cal H}^{int}_{1s}$.
Therefore, in contrast to the case of ${\cal H}^{full-int}_{1s}$, 
it is reasonable to take $\tilde{\cal H}^{int}_{1s}$ as 
an approximation to $\tilde{\cal H}^{full-int}_{1s}$.
In fact, we will show that the replacement 
$\tilde{\cal H}^{full-int}_{1s}$
$\to$
$\tilde{\cal H}^{int}_{1s}$
is a good approximation, 
which agrees with the WIBM and experimental results.

In short, 
one must perform the projection 
onto the subspace ${\sf H}_{1s}$ to get 
a correct effective interaction of $1s$ excitons.
The projection procedure modifies both the form and 
strengths of the effective interaction.
This is essential to justify the WIBM.
In what follows, 
we will derive the effective interaction 
$\tilde{\cal H}^{int}_{1s}$ of $1s$
excitons in a QW from the conventional interacting electron-hole 
Hamiltonian.

\subsection{Model}

We consider the conduction and the heavy-hole bands in a GaAs QW,
which has a direct band gap.
We start from 
the following conventional form 
of the electron-hole Hamiltonian ${\cal H}_{eh}$:
\begin{eqnarray}
&{\cal H}_{eh}&=\sum_{i}\int {\rm d}x
\hat{\psi}^{\dagger}_{i}(x)
\left(-\frac{\nabla^{2}}{2m_{i}}+E_{i}\right)
\hat{\psi}_{i}(x)
+
\sum_{i,i'}\frac{z_{i}z_{i'}}{2}
\nonumber \\
&\times&\int {\rm d}x {\rm d}x'
\hat{\psi}_{i}^{\dagger}(x)
\hat{\psi}_{i'}^{\dagger}(x')
V({\bf r}_{i}-{\bf r}^{\prime}_{i'})
\hat{\psi}_{i'}(x')
\hat{\psi}_{i}(x).
\label{Heh}
\end{eqnarray}
Here, $V({\bf r})$ denotes the Coulomb potential,
which behaves in a QW of width $L$ as $V({\bf r}) \approx e^2/ \epsilon r$ for 
$|{\bf r}| \gtrsim L$, where $\epsilon$ is the static dielectric constant, and 
$V({\bf r}) \approx$ constant for $|{\bf r}| \lesssim L$.
The calculation is simplified by taking 
the limit $L \to 0$ wherever the singularity at ${\bf r} = {\bf 0}$ 
is irrelevant.
In Eq.\ (\ref{Heh}), $\hat{\psi}_{e(h)}(x)$ is the field operator of
an electron (hole), $i=\{e,h\}$, $z_{e(h)}=1\,(-1)$, 
$x\equiv({\bf r}_{e(h)},J^{z}_{e(h)})$, $\int {\rm
d}x\equiv\sum_{J^{z}_{i}}\int {\rm d}^{2}r_{i}$, and similarly for $i'$
and $x'$.  
The index $J^{z}_{e(h)}$ denotes 
the $z$-component of the total angular momentum, which is a good quantum
number, 
when the $z$-axis is taken in the direction normal to the QW
layers.
The $J^{z}_{h}$ is defined as $-1$ times $J^{z}$
of the corresponding  valence band electron.  
In a GaAs QW, $J^{z}_{h}=\pm3/2$ 
for a heavy hole, and $J^{z}_{e}=\pm1/2$ \cite{Haug}.
A photon with $J^{z}_{ph}=+1(-1)$
creates an electron-hole pair with $J^{z}_{e}=-1/2\, (+1/2)$ and 
$J^{z}_{h}=+3/2\,(-3/2)$ to conserve the total angular momentum.

\subsection{Strategy}
Since all states that are excited by photons are electrically neutral,
the discussion is confined to the charge-neutral sector.  
Then the following exciton operator can be defined \cite{Haug-Schmitt-Rink}:
\begin{eqnarray}
&&
b_{{\bf q}\nu S} \equiv \sum_{J^{z}_{e},J^{z}_{h}}
\int {\rm d}^2 r_{e}{\rm d}^2 r_{h}
\frac{1}{\sqrt{\Omega}}
\exp \left(
{i{\bf q} \cdot {m_e {\bf r}_{e}+ m_h {\bf r}_{h} \over M} }
 \right)
\nonumber\\
&& \times
\varphi_{\nu}({\bf r}_{e}-{\bf r}_{h})
\langle S|J^{z}_{e},J^{z}_{h}\rangle
\hat{\psi}_{e}({\bf r}_{e},J^{z}_{e})\hat{\psi}_{h}({\bf r}_{h},J^{z}_{h}).
\label{ex-operator}
\end{eqnarray}
Here, the plane wave corresponds to the center-of-mass motion of an
electron-hole pair, $\varphi_{\nu}({\bf r})$ is a wavefunction for the {\em e-h} relative motion, 
$\langle S|J^{z}_{e},J^{z}_{h}\rangle$ the Clebsch-Gordan (CG)
coefficient,
$\Omega$ the QW area, and $M \equiv m_{e}+m_{h}$.  
In the following, the heavy hole condition $0 < m_e \ll m_h$ is assumed,
and $\mu(\mu')\equiv m_{e(h)}/M$.

From the explicit calculation of the commutation relation for these operators, they
can be treated as bosonic operators when the particle density is very
low.  
This is satisfied under condition (i) of section \ref{conditions}.
When $\varphi_{\nu}({\bf r})$ is the wave function with the quantum number
$\nu (= 1s, 2p_{+}, 2p_{-},\cdots$) of a hydrogen atom in
two-dimension \cite{Haug}, the operator
$b_{{\bf q}\nu S}$ is identified with the bosonic operator for an exciton with the relative motion
index $\nu$.  
Then, the exciton states are labelled by indices ${\bf q}$, $\nu$ and
$S$, where ${\bf q}$ is the momentum of the center-of-mass motion, 
$\nu$
denotes the set of quantum numbers for the relative motion 
$\nu$, and $S$ denotes combinations of
$J^{z}_{e}$ and $J^{z}_{h}$, as shown in Eqs.\,(\ref{cg-tbl2}) and (\ref{cg-tbl1}).
Since $S$ is related to the total angular momentum, 
it is referred to as ``spin'' index in the following.
For transitions from the heavy-hole band to the electron band 
in a GaAs QW, possible changes of the total angular momentum 
are $\Delta J^{z}=\pm 1$ and $\pm 2$.
We here take 
the final states corresponding to $\Delta J^{z}=+1, -1, +2, -2$
as $S=+, -, \alpha,\beta$, respectively.
They are related with $|J^{z}_{e}, J^{z}_{h}\rangle$ as
\begin{equation}
\left(
\begin{array}{c}
|+\rangle \\
|-\rangle \\
|\alpha\rangle \\
|\beta\rangle
\end{array}
\right)
=
\left(
\begin{array}{c}
|-1/2,+3/2\rangle \\
|+1/2,-3/2\rangle \\
|+1/2,+3/2\rangle \\
|-1/2,-3/2\rangle
\end{array}
\right),
\label{cg-tbl2}
\end{equation}
which should be compared with Eq.\,(\ref{cg-tbl1}) of semiconductors
with different band structures.  
Since the dipole transition is associated with $\Delta J^{z}=\pm1$, 
$|+\rangle$ and $|-\rangle$ are dipole active, coupling to circularly polarized light with  
$J^{z}_{ph} = \pm1$, 
whereas $|\alpha\rangle$ and $|\beta\rangle$
are dipole inactive. 
The general form of Eq.\,(\ref{cg-tbl2}) for elliptically polarized light is discussed in Ref.\,\cite{Ciuti}.  

\section{Interaction of excitons before projection}
\label{before}

In this section, the interaction Hamiltonian corresponding to
Eq.\,(\ref{generalHs}) is obtained in order to clarify the difference 
from the
two-body interaction obtained through the projection procedure which is discussed
in the next section.  
For this purpose, we calculate the
scattering amplitude of excitons without any intermediate states.  

\subsection{1s exciton scattering amplitude}
In this subsection, scattering processes which involve 
only $1s$ excitons are considered without the projection procedure, resulting to the
interaction Hamiltonian of $1s$ excitons through a straightforward transformation.  
  
Such scattering processes are schematically shown in Fig.\,1, where the
index ``ex.{\it i}'' should be read as the set of indices $\{{\bf k}_{i},
\nu_{i}, S_{i}\}$.  
In this subsection, $\nu_{i}=1s$ for any $i$ and the index is  dropped if
any confusion is not expected. 
These processes are composed of two parts: one is
a direct process, Fig.\,1\,(a)$\sim$(d) and the other is a fermionic
exchange process, Fig.\,1\,(e)$\sim$(h).  
The form of the interaction Hamiltonian of $1s$ excitons is
\begin{equation}
 {\cal H}^{int}_{1s}=\frac{1}{2\Omega}\sum U({\bf q};\{S\})b^{\dagger}_{{\bf k}+{\bf q}, S_{1}}b^{\dagger}_{{\bf k}'-{\bf q}, S_{2}}
                     b_{{\bf k}', S_{3}}b_{{\bf k}, S_{4}},
\end{equation}
where the scattering amplitude is written as
\begin{equation}
 U({\bf q};\{S\})=U^{o}_{D}({\bf q})U^{s}_{D}(S_{1},S_{2};S_{3},S_{4})
            +U^{o}_{Ex}({\bf q})U^{s}_{Ex}(S_{1},S_{2};S_{3},S_{4}).  
\label{scattering-amp}
\end{equation}
The $U^{o}_{D}({\bf q})U^{s}_{D}(S_{1},S_{2};S_{3},S_{4})$ and
$U^{o}_{Ex}({\bf q})U^{s}_{Ex}(S_{1},S_{2};S_{3},S_{4})$ are the direct and the exchange
scattering amplitudes, respectively.  
The expressions of each component in Eq. (\ref{scattering-amp}) are
\begin{eqnarray}
&&U^{o}_{D}({\bf q})=
\Omega\int {\rm d}{\bf r}_{e}{\rm d}{\bf r}'_{e}{\rm d}{\bf r}_{h}{\rm d}{\bf r}'_{h}
\phi_{{\bf k}+{\bf q}}({\bf r}_{e},{\bf r}_{h})\phi_{{\bf k}'-{\bf q}}({\bf r}'_{e},{\bf r}'_{h})
\nonumber\\
&&\qquad\times
\left\{V({\bf r}_{e}-{\bf r}'_{e})+V({\bf r}_{h}-{\bf r}'_{h})-V({\bf r}_{e}-{\bf r}'_{h})-V({\bf r}'_{e}-{\bf r}_{h})\right\}
\phi_{{\bf k}}({\bf r}_{e},{\bf r}_{h})\phi_{{\bf k}'}({\bf r}'_{e},{\bf r}'_{h}),
\\
&&U^{s}_{D}(S_{1}, S_{2};S_{3}, S_{4})=\sum_{J^{z}_{e},J^{z'}_{e},J^{z}_{h},J^{z}_{h}}
\langle S_{1}|J^{z}_{e},J^{z}_{h}\rangle
\langle S_{2}|J^{z'}_{e},J^{z'}_{h}\rangle
\langle S_{3}|J^{z}_{e},J^{z}_{h}\rangle
\langle S_{4}|J^{z'}_{e},J^{z'}_{h}\rangle,
\\
&&U^{o}_{Ex}({\bf q})=
-\Omega\int {\rm d}{\bf r}_{e}{\rm d}{\bf r}'_{e}{\rm d}{\bf r}_{h}{\rm d}{\bf r}'_{h}
\phi_{{\bf k}+{\bf q}}({\bf r}_{e},{\bf r}_{h})\phi_{{\bf k}'-{\bf q}}({\bf r}'_{e},{\bf r}'_{h})
\nonumber\\
&&\qquad\times
\left\{V({\bf r}_{e}-{\bf r}'_{e})+V({\bf r}_{h}-{\bf r}'_{h})-V({\bf r}_{e}-{\bf r}'_{h})-V({\bf r}'_{e}-{\bf r}_{h})\right\}
\phi_{{\bf k}}({\bf r}'_{e},{\bf r}_{h})\phi_{{\bf k}'}({\bf r}_{e},{\bf r}'_{h}),
\\
{\rm and}\nonumber
\\
&&U^{s}_{Ex}(S_{1}, S_{2};S_{3}, S_{4})=\sum_{J^{z}_{e},J^{z'}_{e},J^{z}_{h},J^{z'}_{h}}
\langle S_{1}|J^{z}_{e},J^{z}_{h}\rangle
\langle S_{2}|J^{z'}_{e},J^{z'}_{h}\rangle
\langle S_{3}|J^{z'}_{e},J^{z}_{h}\rangle
\langle S_{4}|J^{z}_{e},J^{z'}_{h}\rangle,
\end{eqnarray}
where $\phi_{{\bf q}}({\bf r}_{e},{\bf r}_{h})$ is the product of the wave functions of
the center of mass- and the relative motion of excitons,
\begin{equation}
\phi_{{\bf q}}({\bf r}_{e},{\bf r}_{h})\equiv \frac{1}{\sqrt{\Omega}}e^{i{\bf q}\cdot(\mu{\bf r}_{e}+\mu'{\bf r}_{h})}
\varphi({\bf r}_{e}-{\bf r}_{h}).
\end{equation}
The wave function of the relative motion $\varphi$ is the $1s$-wave
function of a hydrogen atom in two-dimension
\begin{equation}
 \varphi({\bf r})=\frac{2\sqrt{2}}{\sqrt{\pi}a_{0}} e^{-2|{\bf r}|/a_{0}},
\end{equation}
where $a_{0}$ is the exciton Bohr radius.  
In the following, $U^{o}_{D(Ex)}$ and $U^{s}_{D(Ex)}$ are referred to as
``orbital'' and ``spin'' parts, respectively.  

First, the orbital parts are calculated.  
Fourier representations of orbital parts in direct and exchange scattering amplitude are
\begin{eqnarray}
U_{D}({\bf q})&=&
\sum_{{\bf p}_{1},{\bf p}_{2}}\tilde{V}\left({{\bf q}}\right)
\biggl[
|\tilde{\varphi}({\bf p}_{1})|^{2}|\tilde{\varphi}({\bf p}_{2})|^{2}
\nonumber\\
&&
+\tilde{\varphi}({\bf p}_{1})\tilde{\varphi}^{*}({\bf p}_{1}+{\bf q})
 \tilde{\varphi}({\bf p}_{2})\tilde{\varphi}^{*}({\bf p}_{2}-{\bf q})
-2\tilde{\varphi}({\bf p}_{1})|\tilde{\varphi}({\bf p}_{2})|^{2}
  \tilde{\varphi}^{*}({\bf p}_{1}+{\bf q})
\biggr],
\label{direct-q}
\\
U_{Ex}({\bf q})&=&
-\sum_{{\bf p}_{1},{\bf p}_{2}}\tilde{V}\left({\bf q}+{\bf p}_{1}-{\bf p}_{2}\right)
\biggl[
|\tilde{\varphi}({\bf p}_{1})|^{2}|\tilde{\varphi}({\bf p}_{2})|^{2}
\nonumber\\
&&
+\tilde{\varphi}({\bf p}_{1})\tilde{\varphi}^{*}({\bf p}_{1}+{\bf q})
 \tilde{\varphi}({\bf p}_{2})\tilde{\varphi}^{*}({\bf p}_{2}-{\bf q})
-2\tilde{\varphi}({\bf p}_{1})|\tilde{\varphi}({\bf p}_{2})|^{2}
  \tilde{\varphi}^{*}({\bf p}_{2}-{\bf q})\biggr].
\label{exchange-q}
\end{eqnarray}
Here, the notations with tilde are defined as
\begin{eqnarray}
\tilde{V}{({\bf p})}&=&\int {\rm d}^{2}{\bf r}e^{i{\bf p}\cdot{\bf r}}\frac{1}{|{\bf r}|}
=\frac{2\pi}{|{\bf p}|}, \\
\tilde{\varphi}({\bf p})&=&
\frac{1}{\sqrt{\Omega}}\int {\rm d}^{2}{\bf r}e^{i{\bf p}\cdot{\bf r}}\varphi({\bf r})
=\frac{1}{\sqrt{\Omega}}\frac{\sqrt{2\pi}a_{0}}{[1+(|{\bf p}|a_{0}/2)^{2}]^{3/2}}.
\end{eqnarray}
Since the transferred momentum in the exciton scattering processes 
is fairly small, which is of the order of the photon momentum,
the direct and the double fermionic exchange interactions are
negligible \cite{Ciuti} and the momentum dependence of the exchange interaction can
be omitted.  
This allows the approximation $U^{o}_{D}({\bf q})\approx U^{o}_{D}({\bf q}=0)\equiv
U^{o}_{D}$ and $U^{o}_{Ex}({\bf q})\approx U^{o}_{Ex}({\bf q}=0)\equiv U^{o}_{Ex}$.    
The ${\bf q}$ dependence of the interaction strength is discussed in
Ref. \cite{Ciuti}.  
Under these condition,  $U^{o}_{D}$ and $U^{o}_{Ex}$ can be directly
obtained from (\ref{direct-q}) and (\ref{exchange-q}) for ${\bf q}\rightarrow 0$:
\begin{eqnarray}
U^{o}_{D}&=&0,
\label{direct}\\
U^{o}_{Ex}&=&2\sum_{{\bf p},{\bf p}'}\tilde{V}({\bf p}-{\bf p}')
\left[|{\tilde\varphi}({\bf p})|^{2}{\tilde\varphi({\bf p})}^{*}{\tilde\varphi}({\bf p}')
-|{\tilde\varphi}({\bf p})|^{2}|{\tilde\varphi}({\bf p}')|^{2}\right].
\label{1-exchange-orbit}
\end{eqnarray}
Equation (\ref{direct}) reflects the charge neutrality of the system.  
\subsection{Scattering amplitude of higher exciton states than $1s$}
The calculation of the exciton scattering amplitude which includes
excitons with $\nu>1s$ is quite similar to the previous calculation
involving excitons with only $\nu=1s$.  
For such scattering processes that satisfy the conservation laws,
the amplitude for scattering from initial state $(\nu_{1}, \nu_{2})$ to the
final state $(\nu_{3},\nu_{4})$ is obtained as
\begin{eqnarray}
 U^{o}_{D \{\nu\}}&=&0,
\\
 U^{o}_{Ex \{\nu\}}&=&2\sum_{{\bf p},{\bf p}'}\tilde{V}({\bf p}-{\bf p}')
\biggl[{\tilde\varphi_{\nu_{1}}({\bf p})}^{*}{\tilde\varphi}_{\nu_{2}}({\bf p})^{*}
{\tilde\varphi}_{\nu_{3}}({\bf p}){\tilde\varphi}_{\nu_{4}}({\bf p}')
\nonumber\\
&&-{\tilde\varphi}_{\nu_{1}}({\bf p})^{*}{\tilde\varphi}_{\nu_{2}}({\bf p}')^{*}
{\tilde\varphi}_{\nu_{3}}({\bf p}){\tilde\varphi}_{\nu_{4}}({\bf p}')\biggr],
\label{Uoexnu}
\end{eqnarray}
under the same assumptions as in the previous calculation.  
Here, ${\tilde\varphi}_{\nu}({\bf p})$ is the Fourier transform of the
corresponding wave function in the real space:
\begin{equation}
 \tilde{\varphi}_{\nu}({\bf p})=
\frac{1}{\sqrt{\Omega}}\int {\rm d}^{2}{\bf r}e^{i{\bf p}\cdot{\bf r}}\varphi_{\nu}({\bf r}).
\end{equation}
\subsection{Interaction Hamiltonian of excitons before projection}
From the results in previous two subsections, the bosonic Hamiltonian of excitons
${\cal H}\equiv{\cal H}^{0}+{\cal H}^{int}$ is obtained
when it is assumed  that the exciton density is low, where ${\cal H}^{0}
$ is the free part of excitons.  
The general form of ${\cal H}^{int}$ through a straightforward transformation is
\begin{equation}
{\cal H}^{int}=\sum_{{\bf k}{\bf k}'{\bf q}\{\nu\}\{S\}}
\frac{U^{o}_{Ex \{\nu\}}U^{s}_{Ex \{S\}}}{2\Omega}
b^{\dagger}_{{\bf k}+{\bf q}\nu_{1} S_{1}}
b^{\dagger}_{{\bf k}'-{\bf q}\nu_{2} S_{2}}
b_{{\bf k}'\nu_{4} S_{4}}
b_{{\bf k}\nu_{3} S_{3}}.
\label{gint}
\end{equation}
Using this formula, we express ${\cal H}$ as
\begin{equation}
{\cal H}
=
{\cal H}^{0}
+
{\cal H}^{\pm}_{1s} 
+ {\cal H}^{\prime}_{1s}
+{\cal H}_{others}, 
\label{Hx}\end{equation}
where ${\cal H}^{\pm}_{1s}+{\cal H}^{\prime}_{1s}\equiv {\cal
H}^{int}_{1s}$, (see Eq.\,(\ref{generalHs})), includes the $\nu =
1s$ operators only, and ${\cal H}_{others}$ denotes such remaining terms
as the interaction between $1s$ and $2p$ excitons, the interaction between
$2p$ and $2p$ excitons and other similar interactions.   
The ${\cal H}^{\pm}_{1s}$ consists
only of the operators with $S=\pm$, corresponding to dipole active excitons, whereas 
${\cal H}^{\prime}_{1s}$ consists of terms
of $S=\alpha$ and $\beta$ operators, including
cross terms with $S=\pm$ operators.  

From the explicit calculation of the spin part of the exchange scattering
amplitude, $U^{s}_{Ex \{S\}}$, for all combinations of
$\{S_{1}, S_{2}, S_{3}, S_{4}\}=\{+, -, \alpha, \beta\}$, 
${\cal H}^{\pm}_{1s}$ and ${\cal H}^{'}_{1s}$ are obtained as
\begin{eqnarray}
{\cal H}^{\pm}_{1s}&=&
\frac{U}{2 \Omega}\sum_{S=\pm} \sum_{{\bf k}{\bf k}'{\bf q}}
b^{\dagger}_{{\bf k}+{\bf q}\,S}b^{\dagger}_{{\bf k}'-{\bf q}\,S}
b^{\phantom{\dagger}}_{{\bf k}'\,S}b^{\phantom{\dagger}}_{{\bf k}\,S},
\label{Hexpm}
\\
{\cal H}^{\prime}_{1s}&=&
{U \over \Omega} \sum_{{\bf k}{\bf k}'{\bf q}} 
\Biggl[
\sum_{S = \alpha, \beta} 
\biggl(
\frac{1}{2} b^{\dagger}_{{\bf k}+{\bf q}\,S}b^{\dagger}_{{\bf k}'-{\bf q}\,S}
b^{\phantom{\dagger}}_{{\bf k}'\,S}b^{\phantom{\dagger}}_{{\bf k}\,S}
\nonumber\\ 
&&+ 
b^{\dagger}_{{\bf k}+{\bf q}\,+}b^{\dagger}_{{\bf k}'-{\bf q}\,S}
b^{\phantom{\dagger}}_{{\bf k}'\,S}b^{\phantom{\dagger}}_{{\bf k}\,+}
+
b^{\dagger}_{{\bf k}+{\bf q}\,-}b^{\dagger}_{{\bf k}'-{\bf q}\,S}
b^{\phantom{\dagger}}_{{\bf k}'\,S}b^{\phantom{\dagger}}_{{\bf k}\,-}
\biggr)
\nonumber\\
&&+
\biggl(
b^{\dagger}_{{\bf k}+{\bf q}\,+}b^{\dagger}_{{\bf k}'-{\bf q}\,-}
b^{\phantom{\dagger}}_{{\bf k}'\,\alpha}b^{\phantom{\dagger}}_{{\bf k}\,\beta}
+ {\rm h.c.}
\biggr)
\Biggr],
\label{Hex'}
\end{eqnarray}
where 
$b_{{\bf q} S}\equiv b_{{\bf q} 1s S} $,
and the effective 
interaction strength $U\equiv U^{o}_{Ex \{\nu\}=\{1s\}}$.

For $d=2$, and in the limit of $L \to 0$, 
Eq.\ (\ref{1-exchange-orbit}) is evaluated as
\begin{equation}
U
=2\pi a^{2}_{0}
\left(1-\frac{315\pi^{2}}{4096}\right)E^{b}_{ex}
\approx 1.52a^{2}_{0}E^{2D}_{ex},
\end{equation}
where $a_{0}$ is the exciton Bohr radius
and $E^{2D}_{ex}$ is the binding energy of the two-dimensional exciton
\cite{Schmitt-Rink2}. 
Note that the estimation of Eq.\,(\ref{1-exchange-orbit}) in the
three-dimensional case, using three-dimensional $1s$ wave function,
gives the hard-core scattering strength
obtained in Ref.\,\cite{Hanamura}.  
This fact shows that the theory presented  in this section corresponds to
the result obtained by the Usui transformation \cite{Usui}.  

The point which should be emphasized in the dipole active part is that
there is no interaction
terms between the exciton with $S=+$ and the exciton with $S=-$
(opposite spin exciton interaction), which is represented in the form of
$
\propto b^{\dagger}_{+}b^{\dagger}_{-}b_{-}b_{+}
$, and that there are only the interaction terms between the excitons
with $S=+(-)$ (equal spin exciton interaction).
This is due to vanishing of $U^{s}_{Ex \{S\}}$ for the corresponding combination
of $\{S\}$.  
Here two ways of thinking are possible.  
The one is that the excitons with $S=+$ and the excitons with $S=-1$
will not interact with each other.  The other is that the absence of such an
interaction term is due to the fact that so far we calculated to lowest
order only.  
A large number of experiments show that excitons with the opposite
spins do interact and that such an interaction is crucial,
for instance, for four-wave mixing in the time-domain.  
So it is not appropriate to interpret ${\cal H}^{\pm}_{1s}$ as the effective
Hamiltonian for dipole active $1s$ excitons.  
The straightforward transformation presented
in this section is not appropriate for deriving the effective
Hamiltonian of $1s$ excitons and
the projection procedure discussed in the next section is indispensable.

\section{Projection procedure}
\label{projection}
In this section, the interaction Hamiltonian of $1s$ excitons,
which corresponds to Eq.\,(\ref{projectHs}),
is obtained through the projection procedure \cite{Inoue}. 
This procedure yields the correct two-body interactions of 
$1s$ excitons in the
subspace of $1s$ excitons.  
In particular, 
the interaction term of excitons with
the opposite spins is obtained through the projection, 
whereas such an interaction was not obtained in the 
previous section.
In other words, such an interaction 
is obtained when going beyond the HF approximation of $1s$ excitons
alone. 
Moreover, the projection yields a large renormalization of 
the interaction strength of excitons with
the same spins.  
Among works which have the similar motivation of going beyond HF, 
modification of the
exciton binding energy beyond HF is
discussed by considering screening effects in Ref.\,\cite{Fernandez}.  

In the previous section, the effects of higher exciton states have been
completely excluded in deriving the interaction Hamiltonian of $1s$ excitons,
Eqs.\,(\ref{Hexpm}) and (\ref{Hex'}).  
In this section, the theory is discussed in the subspace spanned by
$1s$ excitons, where the effects of higher exciton states are
renormalized.  
Since this scenario is quite similar to the derivation of an effective
Hamiltonian from the Hubbard model with large on-site Coulomb repulsion
through a projection procedure \cite{Emery}, the method used here is
referred to as projection.  
In other words, this is nothing but the real part of the second order vertex correction
in the filed theory, yielding the energy shift.  

The higher exciton states are taken as the intermediate states in the
scattering processes of $1s$ excitons.  
Schematically, the scattering processes of excitons shown in Fig.\,2 are
considered.  
Since our purpose is to obtain the effective interaction of $1s$
excitons, the relative motion indices of four external lines must be
$\nu=1s$.  
As for the intermediate states, excitons which are connected with $1s$
excitons by dipole transitions are considered.  
Then, the lowest energy excitons for the intermediate states are
$2p_{\pm}$ excitons.  
Note that in two-dimensional system, the $p$ states are doubly degenerated.  

The Hamiltonian for the $1s-\nu$ interaction processes is
\begin{eqnarray}
{\cal H}_{\rm out}&\equiv&\sum_{\nu} {\cal H}_{1s-\nu}
\nonumber\\
&=& \sum_{\nu}\sum_{\{S\}}\sum_{{\bf k}{\bf k}'{\bf q}}g_{\nu}({\bf q})
b^{\dagger}_{1s,{\bf k}+{\bf q},S_{1}}b^{\dagger}_{1s,{\bf k}'-{\bf q},S_{2}}
b_{\nu,{\bf k}',S_{4}}b_{\nu,{\bf k},S_{3}}+{\rm h.c.}.  
\end{eqnarray}
Let ${\cal H}_{\rm org}$ be the Hamiltonian which includes the kinetic
and the interaction terms of $1s$ and $\nu$ excitons, and with
eigenstates and eigenenergies defined by
\begin{equation}
 {\cal H}_{\rm org}|\Phi\rangle=E_{\Phi}|\Phi\rangle.
\end{equation}
Consider the following Schr\"odinger equation:
\begin{equation}
 \left({\cal H}_{\rm org}+{\cal H}_{\rm out}\right)|\Psi\rangle=E|\Psi\rangle.
\end{equation}
Symbolically, the solution of this equation is shown as
\begin{eqnarray}
 |\Psi\rangle&=&\frac{{\cal H}_{\rm out}}{E-{\cal H}_{\rm org}}|\Psi\rangle
\nonumber\\
&=&\sum_{\Phi}|\Phi\rangle\frac{\langle\Phi|{\cal H}_{\rm out}|\Psi\rangle}{E-E_{\Phi}}
+\frac{{\cal P}}{E-{\cal H}_{\rm org}}{\cal H}_{\rm out}|\Psi\rangle,
\end{eqnarray}
where the projection operator ${\cal P}$ is defined as
\begin{equation}
 {\cal P}\equiv 1-\sum_{\Phi}|\Phi\rangle\langle\Phi|.
\end{equation}
The $|\Psi\rangle$ is rewritten as
\begin{equation}
 |\Psi\rangle=\sum_{\Phi}a_{\Phi}|\Psi_{\Phi}\rangle,
\end{equation}
where
\begin{equation}
 a_{\Phi}\equiv \frac{\langle\Phi|{\cal H}_{\rm out}|\Psi\rangle}{E-E_{\Phi}},
\end{equation}
and
\begin{equation}
 |\Psi_{\Phi}\rangle\equiv |\Phi\rangle+
\frac{{\cal P}}{E-{\cal H}_{\rm org}}{\cal H}_{\rm out}|\Psi_{\Phi}\rangle.
\end{equation}
In lowest order, this wave function can be approximated as
\begin{equation}
 |\Psi_{\Phi}\rangle
\simeq|\Phi\rangle+\frac{1}{E-{\cal H}_{\rm org}}{\cal H}_{\rm out}|\Phi\rangle.
\end{equation}
From them, the Schr\"odinger equation for the projected out Hamiltonian,
that is, the effective Hamiltonian is obtained as
\begin{equation}
 \left(E-E_{\Phi}\right)a_{\Phi}=
\sum_{\Phi'}a_{\Phi'}
\langle\Phi|{\cal H}_{\rm out}\frac{1}{E-{\cal H}_{org}}{\cal H}_{\rm out}|\Phi'\rangle.
\end{equation}

Here, two successive exchange scattering processes are considered, whose individual
scattering amplitudes are obtained from Eq.\,(\ref{Uoexnu}).  
The resolvent $1/(E-{\cal H}_{org})$ is replaced with the difference of
the kinetic energy between two
$\nu$ excitons and two $1s$ excitons, because the interaction energy of
each exciton can be assumed to be small as compared to their kinetic
energy.  
The orbital part of the renormalized scattering amplitude of $1s$
excitons, ${U^{o \prime}_{Ex}}$, which is
calculated by second order perturbative calculation, is
\begin{eqnarray}
{U^{o \prime}_{Ex}}
&=&
\frac{1}{\Omega}
\sum_{{\bf K},\nu\neq 1s}
\frac{|g_{\nu}({\bf K})|^{2}}{2\left(E_{\nu}+{\bf K}^2/2M\right)-2E_{1s}}
\nonumber\\
&=&
\frac{1}{\Omega}
\sum_{{\bf K},\nu\neq 1s}
\frac{1}{2\left(E_{\nu}+{\bf K}^2/2M\right)-2E_{1s}}
\Biggl|
\sum_{{\bf p},{\bf p}'}\tilde{V}({\bf p}-{\bf p}'+{\bf K})
\Bigl[
-\tilde{\varphi}^{*}_{1s}({\bf p})\tilde{\varphi}^{*}_{1s}({\bf p}')
\tilde{\varphi}_{\nu}({\bf p})\tilde{\varphi}_{\nu}({\bf p}')
\nonumber\\
&&
+2\tilde{\varphi}^{*}_{1s}({\bf p})\tilde{\varphi}^{*}_{1s}({\bf p}-{\bf K})
\tilde{\varphi}_{\nu}({\bf p})\tilde{\varphi}_{\nu}({\bf p}')
-\tilde{\varphi}^{*}_{1s}({\bf p})\tilde{\varphi}^{*}_{1s}({\bf p}')
\tilde{\varphi}_{\nu}({\bf p}-{\bf K})\tilde{\varphi}_{\nu}({\bf p}'+{\bf K})
\Bigr]
\Biggr|^{2}. 
\label{U'}
\end{eqnarray}
As for the spin part, ${U^{s\prime}_{Ex}}$, it is obtained by the simple
product of two successive spin weights $U^{s}_{Ex}$, 
\begin{equation}
{U^{s\prime}_{Ex}}(S_{1}, S_{2};S_{3}, S_{4})
=\sum_{S,S'}U^{s}_{Ex}(S_{1}, S_{2};S, S')U^{s}_{Ex}(S, S';S_{3}, S_{4}).
\end{equation}
From the explicit representation of the renormalized scattering
amplitude $U'\equiv{U^{o\prime}_{Ex}}{U^{s\prime}_{Ex}}$, the renormalized
Hamiltonian of $1s$ excitons is written in the form
\begin{equation}
{\tilde{\cal H}}_{1s}
=
{\tilde{\cal H}}^{0}_{1s}
+
{\tilde{\cal H}}^\pm_{1s}+{\tilde{\cal H}}'_{1s},
\label{tildeHex}
\end{equation}
where
${\tilde{\cal H}}^{0}_{1s}$ is the Hamiltonian of free $1s$ excitons, 
and
${\tilde{\cal H}}^{\pm}_{1s}$+${\tilde{\cal H}}'_{1s}\equiv\tilde{\cal H}^{int}_{1s}$
, (see Eq.\,(\ref{projectHs})), includes the $\nu = 1s$ operators only.  
The ${\tilde{\cal H}}^{\pm}_{1s}$ consists
only of operators with $S=\pm$, whereas
${\tilde{\cal H}}^{\prime}_{1s}$ consists of terms
of $S=\alpha$ and $\beta$ operators, including
cross terms with $S=\pm$ operators. 

The interaction Hamiltonian which has only dipole active $1s$ excitons  
${\tilde{\cal H}}^\pm_{1s}$, is
\begin{eqnarray}
{\tilde{\cal H}}^\pm_{1s}&=&\frac{U-U'}{2 \Omega}
\sum_{S=\pm}\sum_{{\bf k}{\bf k}'{\bf q}}
b^{\dagger}_{{\bf k}+{\bf q}\,S}b^{\dagger}_{{\bf k}'-{\bf q}\,S}
b^{\phantom{\dagger}}_{{\bf k}'\,S}b^{\phantom{\dagger}}_{{\bf k}\,S}
\nonumber\\
&&
- {U' \over \Omega}
\sum_{{\bf k}{\bf k}'{\bf q}}
b^{\dagger}_{{\bf k}+{\bf q}\,+}b^{\dagger}_{{\bf k}'-{\bf q}\,-}
b^{\phantom{\dagger}}_{{\bf k}'\,-}b^{\phantom{\dagger}}_{{\bf k}\,+},
\label{tildeHexpm}
\end{eqnarray}
where $U'={U^{o}_{Ex}}'$ is a positive constant
which arises from the renormalization of 
higher exciton states
$(\nu=2p_{+},\,2p_{-},\,\cdots)$.  
Comparing the right-hand side of Eq.\ (\ref{tildeHexpm}) 
with that of Eq.\ (\ref{Hexpm}), 
we see that 
the coefficient of the first term is renormalized as 
$U \to U-U'$, and that the second term is generated through the
projection procedure, which leads to the opposite spin exciton
interaction.  
This is due to the fact that though the spin weight
$U^{s}_{Ex}(+,-;+,-)$ vanishes, the
spin weight ${U^{s}_{Ex}}'(+,-,;+,-)\neq0$ because the intermediate
states can make use of the state with the $S=\{\alpha, \beta\}$, and
$U^{s}_{Ex}(+,-;\alpha,\beta)\neq0$.  
This shows that
the renormalization of higher exciton states results in 
a renormalized Hamiltonian ${\tilde{\cal H}}^\pm_{1s}$,
which differs, both quantitatively and qualitatively, 
 from the Hamiltonian ${\cal H}^{\pm}_{1s}$, where higher exciton states
 than $1s$ exciton are completely ignored.

The renormalized interaction Hamiltonian which includes dipole inactive
excitons is obtained similarly as
\begin{eqnarray}
 {\tilde{\cal H}}'_{1s}&=&\frac{U-U'}{\Omega}\sum_{S=\alpha,\beta}\sum_{{\bf k}{\bf k}'{\bf q}}
\biggl[
\frac{1}{2}b^{\dagger}_{{\bf k}+{\bf q}\,S}b^{\dagger}_{{\bf k}'-{\bf q}\,S}b_{{\bf k}'\,S}b_{{\bf k}\,S}
\nonumber \\
&&\qquad+b^{\dagger}_{{\bf k}+{\bf q}\,+}b^{\dagger}_{{\bf k}'-{\bf q}\,S}b_{{\bf k}'\,S}b_{{\bf k}\,+}
  +b^{\dagger}_{{\bf k}+{\bf q}\,-}b^{\dagger}_{{\bf k}'-{\bf q}\,S}b_{{\bf k}'\,S}b_{{\bf k}\,-}\biggr]
\nonumber \\
&&+\frac{U}{\Omega}\sum_{{\bf k}{\bf k}'{\bf q}}
\biggl(
b^{\dagger}_{{\bf k}+{\bf q}\,\alpha}b^{\dagger}_{{\bf k}'-{\bf q}\,\beta}b_{{\bf k}'\,\beta}b_{{\bf k}\,\alpha}
+{\rm h.c.}
\biggr)
\nonumber\\
&&-\frac{U'}{\Omega}\sum_{{\bf k}{\bf k}'{\bf q}}
b^{\dagger}_{{\bf k}+{\bf q}\,\alpha}b^{\dagger}_{{\bf k}'-{\bf q}\,\beta}b_{{\bf k}'\,\beta}b_{{\bf k}\,\alpha}.
\label{tildeHex'}
\end{eqnarray}
Since ${\tilde{\cal H}}'_{1s}$ includes 
dipole inactive $1s$ excitons with $S=\alpha$, $\beta$,
it does not contribute to the optical response 
in its lowest order \cite{lowestorder,inactive}.  

\section{Discussions and Remarks}
\subsection{Microscopic foundation of the WIBM}
Kuwata-Gonokami {\em et al.} \cite{Kuwata-Gonokami} introduced a
phenomenological Hamiltonian, WIBM, which
yields good agreement with the experimental four-wave mixing data.  
The WIBM has two kinds of interaction terms of excitons: 
One is a repulsive term $R$ for the same spin excitons.
The other is an attractive interaction $W$ for the opposite spin excitons.  
As our first important result, we note that the phenomenological
Hamiltonian \cite{Kuwata-Gonokami}
has the same form as our ${\tilde{\cal H}}_{1s}^{\pm}$, 
the dipole active part of $\tilde{\cal H}_{1s}$.
This is quite reasonable because the other part ${\tilde{\cal H}}'_{1s}$, 
which is dipole inactive, 
should be invisible in low-order optical experiments \cite{inactive}.
We can, therefore, identify the parameters $R$ and $W$ of the
phenomenological Hamiltonian
\cite{Kuwata-Gonokami} as
\begin{eqnarray}
R&=&\frac{U-U'}{2\Omega},
\label{R}
\\
W &=& -\frac{U'}{\Omega}.
\label{W}\end{eqnarray}
The value of $U'$, as given by Eq.\ (\ref{U'}),  
depends on the material parameters such as 
$M$ and $\epsilon$, and hence is different for different 
materials.
It also depends on the QW parameter $L$.
Moreover, when imperfections in the QW are non-negligible, 
the expressions of $U'$ should be modified accordingly.
Therefore, even for the same material the values of $R$ and $W$ could
vary from sample to sample,
which seems to be consistent with recent experimental results \cite{Kuwata-Gonokami1}.
Note, however, that the {\em existence} of both interaction terms of
excitons ${\tilde{\cal H}
}^{\pm}_{1s}$ is independent of such details.  

Since the accurate evaluation of Eq.\ (\ref{U'}) is rather 
tedious, we here estimate the typical value of $U'$ as follows.
The ${\bf K}$-summation in Eq.\ (\ref{U'}) is cut off 
for $K \gtrsim C_L / L$ (through the cutoff of ${\tilde V}$) 
and/or for $K \gtrsim C_{a_0} / a_0$ (through $\tilde \varphi_\nu$),
where $C_L$ and $C_{a_0}$ are cutoff parameters of the order of unity.
For the case of the QW sample of 
Ref.\,\cite{Kuwata-Gonokami}, 
$L \approx a_0$, hence we may cutoff 
the ${\bf K}$-summation 
for $K \gtrsim C / a_0$, where $C$ is of the order of unity.
For the $\nu$ summation, 
we may consider
$\nu=2p_{\pm}$ states only, because higher exciton states 
give much smaller overlap integrals.
The summations for ${\bf p}$ and ${\bf p}'$ are replaced with integrals and $U'$ is
evaluated, as shown in appendix B, as
$U'\approx 16.5a^{2}_{0}C^{2}E^{2D}_{ex}$.

Reference\,\ \cite{Kuwata-Gonokami} reported the ratio
$R:W$ as $1:-15$.
From Eqs.\ (\ref{R}) and (\ref{W}), 
we find that this ratio is reproduced by the present theory
when the cutoff parameter $C \sim 0.3$, 
which is consistent with the requirement that $C$ is of the order of unity.
Considering that the values of $R$ and $W$ vary slightly from sample
to sample \cite{Kuwata-Gonokami1}, the agreement seems satisfactory.
Note that such a small value of $R$ reported in
Ref.\,\cite{Kuwata-Gonokami} is due to the renormalization of $U\to U-U'$.  
Once the agreement of ${\tilde{\cal H}}_{1s}^{\pm}$ 
with the phenomenological Hamiltonian is established, 
the agreement with the experiment follows, as presented in
Refs.\,\cite{Kuwata-Gonokami,Shirane,Svirko}.
That is, 
lowest-order perturbative calculations
for the polariton-polariton scattering amplitudes  
agree with the experiment \cite{Kuwata-Gonokami,lowestorder}.
From the above discussion, we conclude that the present theory yields the
microscopic background of the two exciton interaction terms in the WIBM.

\subsection{Correct effective Hamiltonian of $1s$ excitons}
The correct form of the effective Hamiltonian of $1s$ excitons is 
the renormalized one instead of the Hamiltonian which is obtained
without the projection procedure, i.e.,   
\begin{equation}
{\cal H}^{eff}_{1s} = {\tilde{\cal H}}^\pm_{1s} + {\tilde{\cal
H}}'_{1s}.   
\end{equation}
In fact, the opposite spin exciton interaction in ${\tilde{\cal H}}_{1s}^{\pm}$, 
which is absent in ${\cal H}_{1s}^{\pm}$, 
has been clearly observed experimentally in Refs.\ 
\cite{Bott,Kuwata-Gonokami}.
We have used a low-order perturbation theory 
to derive $\tilde{\cal H}_{1s}$, where successive scatterings in the
intermediate processes are not considered.  
However, this does not imply a total neglect of
multi scattering processes, 
because we have calculated a Hamiltonian rather than observables.
In fact, 
a systematic summing up of higher order scattering processes is incorporated in our theory 
if one calculates 
higher-order perturbation terms from $\tilde{\cal H}_{1s}$ by writing
down the Bethe-Salpeter
equation.  

Note that 
${\tilde{\cal H}}_{1s}$ is not positive definite to the fourth
 order in the exciton operators.
The stability of the system should be preserved by higher order terms.
In general situations, 
properties of a system described by such a Hamiltonian 
should not be analyzed by a perturbation theory based on 
the vacuum of the free part. 
Nevertheless, we can use such a perturbation theory 
in our case, because our 
exciton theory has the built-in constraint that the ground state 
is the state with no excitons, i.e., the vacuum of ${\tilde{\cal H}}^{0}_{1s}$.
The effective Hamiltonian ${\tilde{\cal H}}_{1s}$ together with this constraint 
constitutes a consistent theory, which justifies 
the low-order perturbation theory based on the given vacuum, if
the optical excitation is sufficiently weak.

\subsection{Comparison with existing bosonic theories}
As mentioned in the introduction, most of the existing bosonic theories
treated spinless excitons and
argued that the effect of spins is obtained by trivial extension of the
spinless theory \cite{Hanamura}.  
The result of such theories is that excitons interact
repulsively.  
The interaction of excitons with opposite spins then is attributed to
the biexciton formation.  
However, an explicit expression of an interaction Hamiltonian of
$1s$ excitons with opposite spins was not obtained \cite{Ivanov2}.  
This is quite natural because the conventional method
was discussed in the HF approximation of $1s$
excitons only, where higher
exciton states were completely ignored.  
Such a theory will not yield the interaction of excitons with opposite spins, as shown in this paper.  
The projection method discussed in this paper proves that
one can obtain the explicit form of the interaction Hamiltonian of
excitons with opposite spins beyond HF approximation of only $1s$ excitons.  
There are several papers where such an approach is motivated \cite{Fernandez};
however, only the modification of the exciton binding energy is discussed.  
Using our interaction Hamiltonian, one can reproduce these results \cite{Fernandez2}.

Several authors obtain an interaction between
$S=+$ and $S=-$ excitons within the HF
approximation of only $1s$ excitons \cite{Ivanov2}.  
In those approaches, exciton operators do not have a spin coordinate, and the
exchange interaction is not the correct one in the sense that the spin
coordinate is not exchanged because of the absence of such a coordinate.  
From this theory, the interaction strength of the excitons with opposite spins
is $-1$ times the one with the identical spins.  
The absence of the opposite spin exciton interaction within HF of only
$1s$ excitons is
confirmed in many papers \cite{Fernandez2}.

\subsection{Validity of the theory}
\label{sec_validity}

In deriving the effective Hamiltonian of $1s$ excitons, 
we have assumed three conditions listed in Sec.\,\ref{conditions}:
(i) the excitation is weak 
so that 
the mean distance of photo-created excitons is much
larger than the Bohr radius of the $1s$ exciton,
and 
(ii) all the photon energies (pump, probe, and signal) 
are close to
the energy of the $1s$ exciton,
and
(iii)
the line width of $1s$ exciton is smaller than the 
detuning energies.

The condition (i) allows us to use the 
boson representation of excitons.  
This condition may be confirmed experimentally by the fact that 
the signal intensity is precisely proportional to the square of the pump
intensity, i.e., the optical response is well described by 
$\chi^{(3)}$.
This suggests 
that the two-body scattering processes would be dominant.
When $l_{ex}$ is increased to $l_{ex}\sim a_{0}$,
the deviation from the boson statistics of the operators $b_{{\bf q}\nu S}$
and $b^{\dagger}_{{\bf q}\nu S}$ 
will become non-negligible,  
which invalidates the boson representation \cite{Hawton}.

The condition (ii) means that 
the $1s$ excitons give dominant contributions. 
This allows us to project out all states higher than $1s$.

The final condition (iii) 
allows us to neglect relaxation process of $1s$ excitons.
Namely, 
the equation of motion of 
the reduced density operator $\tilde\rho_{1s}$ 
in the $1s$ exciton subspace generally takes the following form;
\begin{equation}
\frac{\partial \tilde\rho_{1s}}{\partial t}
=\frac{1}{i}[{\tilde{\cal H}}_{1s}, \tilde\rho_{1s}]
+ \tilde\Gamma \tilde\rho_{1s}.
\end{equation}
Here, ${\tilde{\cal H}}_{1s}$ describes the unitary evolution
of $\tilde\rho_{1s}$, and $\tilde\Gamma$ is the relaxation operator,
which
is described by the imaginary part of a vertex correction.  
Since the relaxation processes are less
crucial under the condition (iii), one 
may disregard $\tilde\Gamma$, i.e., one may consider only the real 
part of the vertex correction.
One might think that under 
these conditions the optical signals would not be strong enough 
to obtain experimental data.
However, Kuwata-Gonokami {\em et al.} \cite{Kuwata-Gonokami} 
proposed a genius method to overcome this difficulty: 
They confined a GaAs QW in a high-Q optical cavity.
This results in a large splitting of 
excitonic polariton spectrum.
Some nonlinear optical signals are strongest at the polariton 
energies (upper and lower ones) because of 
the polariton resonance.
On the other hand, 
the dissipation (which 
creates real excitons in the QW) is weak at these energies.
[This may be understood by considering possible final states:
When the initial state is a single photon state (coming from 
an external light source) that 
has the energy of the lower-polariton peak, 
the final state cannot be an exciton state (without a photon)
because the energy is short to create a real exciton.]
Therefore, 
one can obtain strong signals without significant dissipation 
when the photon energies are close to the polariton energies 
in a high-Q optical cavity.
Using this idea, 
Kuwata-Gonokami {\em et al.} \cite{Kuwata-Gonokami}
measured nonlinear optical responses 
under the conditions (i)-(iii), 
and showed the validity of the WIBM.
Our theory is valid in such a case.

On the other hand, Shirane {\it et al.} recently demonstrated that 
the dissipation becomes important when 
the Q value of the optical cavity is lowered
\cite{Shirane}.  
In this case the relaxation processes of $1s$ exciton becomes important, 
which means that
$\tilde{\Gamma}$ must be fully considered. 
These are closely related to the excitation induced relaxation (EID)
discussed below.    
The relaxation effect is one of the future problems.  

Another example to which the present theory is applicable 
may be the optical
Stark effect in high-quality samples at low temperature.  
In the past experiments of the optical Stark effect, 
one had to take the detuning rather large (hence, condition (ii) 
is not satisfied) in order to avoid the absorption tail.
The absorption tail would be reduced for 
samples with better quality and at lower temperatures.

\subsection{Biexciton formation and attractive interaction of excitons}
It has been conjectured \cite{Kuwata-Gonokami} that 
a ``biexciton effect'' would be 
the origin of the ``{\em W} term'', i.e., 
the opposite spin exciton interaction.  
However, this argument is misleading.
The biexciton state is analogous to a hydrogen molecule and is formed 
essentially through the mixing of two $1s$ states having different centers.
The mixing of two hydrogen atoms yields the bonding and antibonding 
states, which are represented as
$
(1/\sqrt{2})
(c^{\dagger}_{1\uparrow}c^{\dagger}_{2\downarrow}
\pm
c^{\dagger}_{1\downarrow}c^{\dagger}_{2\uparrow})
h^{\dagger}_{1\sigma}h^{\dagger}_{2\sigma'}|0\rangle
$.
Here, $c^\dagger_{1 (2)}$ creates an electron in the $1s$ state
located at nucleus 1(2), 
and  $h^\dagger_{1(2)}$ creates the nucleus.  
The lower energy state is the bonding, that is, molecular state.  
In the case of excitons with $J^{z}_{e}=\pm1/2$ and $J_h^{z} = \pm 3/2$,
the corresponding states are
$
(1/\sqrt{2})[
b^{\dagger}_{+}b^{\dagger}_{-}
\pm
b^{\dagger}_{\alpha}b^{\dagger}_{\beta}
]|0\rangle
$,
where the ${\bf k}$-dependence is omitted in order to focus on 
the $S$-dependence.
The bonding state ($-$ sign for a positive coupling constant) has a
lower energy and is called biexciton. 
This energy splitting 
between the bonding and antibonding states
is induced by an interaction of the form of 
$b_+^\dagger b_-^\dagger b_\alpha b_\beta + {\rm h.c.}$, 
which is included in 
$\tilde{\cal H}'_{1s}$  (or, before the renormalization, in 
${\cal H}^{\prime}_{1s}$ of Eq.\ (\ref{Hex'})).
From a more general point of view, a typical bound state in an
interacting boson model is the eigenstate
\begin{equation}
|{\rm bound}\rangle=(a^{\dagger}b^{\dagger}-{\rm sign}(g) c^{\dagger}d^{\dagger})|0\rangle,
\end{equation}
of a boson Hamiltonian in the following form:
\begin{equation}
{\cal H}=\omega(a^{\dagger}a+b^{\dagger}b+c^{\dagger}c+d^{\dagger}d)+
g(a^{\dagger}b^{\dagger}cd+d^{\dagger}c^{\dagger}ba).
\end{equation}
The eigenenergy of $|{\rm bound}\rangle$ is $2\omega-|g|$.
Note that the existence of the bound state is independent of the sign of
	   the interaction $g$.

On the other hand, 
the $W$ term lowers the energies of {\em both} states
{\em by the same amount}, hence 
does not play a central role
in the formation of the biexciton state.
The most important effect of the $W$ term is 
to lower the energy of 
$b^{\dagger}_{+}b^{\dagger}_{-}|0\rangle$, relative to those of 
$b^{\dagger}_{+}b^{\dagger}_{+}|0\rangle$ and
$b^{\dagger}_{-}b^{\dagger}_{-}|0\rangle$, 
and this effect was detected experimentally \cite{Kuwata-Gonokami}.  
In the framework of the present bosonic theory, 
$\tilde{\cal H}'_{1s}$ lowers the energy of the bonding (biexciton)
state relative to that of the antibonding state, and thus is crucial for
the formation of the biexciton state,  whereas the ${\em W}$
term lowers the energy of both bonding and antibonding states.

\subsection{Dipole decoupling and HF approximation} 
The original SBE are a theory within HF.  
There are two reasons that an extension beyond HF approximation becomes
necessary.  
One reason is that it is necessary to take exciton-exciton correlation effects into account.  
When the excitation is low, the optical response from semiconductors are
attributed to excitons.  
Similar features appear under high magnetic fields \cite{Chemla},
where the most striking feature is that the signal of time-domain
four-wave mixing does not decay exponentially, whereas the SBE
predict a single-exponential decay.
These typical two cases are beyond the scope of SBE.

As for the spin degrees of freedom,
the excitations created by photons with the opposite circularly
polarization are completely decoupled within HF.  
Furthermore, the experimental results of the polarization dependence of four-wave
mixing signals and quantum beats are not treated within HF.  
The coupling of the excitation with the opposite spin is obtained beyond
HF approximation.  

Finally, we discuss 
the relation between the fermionic theories 
\cite{Haug,Lindberg,Schafer,Hu,Rappen} and the present bosonic theory.  
The HF factorization treatment
of the SBE \cite{Lindberg}
can not produce the interaction 
between the excitation created by right-circularly polarized light and
the excitation by left-circularly polarized light.
The HF theory, therefore, corresponds to
${\cal H}^{\pm}_{1s}$, Eq.\ (\ref{Hexpm}).
It was argued in Refs.\ \cite{Schafer,Hu,Rappen} that
the interactions of an exciton with higher states 
(including free carriers) 
are important, and that
the interactions result in
the energy shift, the EID, and the ``biexcitonic correlations''.
In the bosonic theory in the form of 
Eq.\ (\ref{Hx}), these effects are included in 
${\cal H}^{\prime}_{1s}$
and 
${\cal H}_{others}$.
After the projection is made,
the relation is roughly as follows.
The renormalized Hamiltonian
${\tilde{\cal H}}^{\pm}_{1s}$, Eq.\ (\ref{tildeHexpm}),
would include the HF term.
The EID may be described by both
$\tilde\Gamma$ and ${\tilde{\cal H}}'_{1s}$.
The ``biexcitonic correlation'' would be included in ${\tilde{\cal H}}'_{1s}$.
We believe that the present theory thus helps to bridge the gap between
the bosonic theories \cite{Hanamura,Axt,Ivanov,Kuwata-Gonokami} 
and the fermionic theories \cite{Haug,Lindberg,Schafer,Hu,Rappen}
of {\em e-h} systems.
However, more detailed comparisons will be a subject of future studies.

As another future problem, the microscopic expression of the filling
factor $\nu$ should be discussed on the same footing.  
Although such an expression was derived in \cite{Hiroshima}, it corresponds to the
``before projection'' in our theory, so that the corresponding term ``after projection'' 
remains a subject of future research.  

\section{Conclusions}
In this paper, we have derived the effective Hamiltonian for $1s$
excitons with spin degree of freedom in two-dimension. 
This theory is valid when excitation density is weak and when the photon
energy is close to the $1s$ exciton energy, because the boson operators
of $1s$ excitons are used.  
Relaxation processes of excitons should be less important like, e.g., a QW in micro cavity
with high-Q value.  
What should be most emphasized is that the projection is crucial to
obtain the effective Hamiltonian of excitons:
the correct effective Hamiltonian of
$1s$ excitons is not the one which is obtained by discarding the all exciton operators of
$\nu\neq1s$ among the full interaction of excitons (see
Eqs.\,(\ref{generalH}) and (\ref{generalHs})), because it can not explain the
experimental results even qualitatively, as discussed in Sec.\,\ref{before}.  
The higher exciton states $\nu=2p, 3d, \cdots$ play important roles as
intermediate states.  
In order to include such effects, the projection is used.  
Through this procedure, the interactions of excitons
with the opposite spins are obtained and the interaction strength of excitons with the same spins
is drastically modified (renormalized) as shown in Sec.\,\ref{projection}.  
In short, the procedure renormalizes both the form and strength of the effective
interaction.
  
It is also shown that the effective Hamiltonian obtained through the
projection provides the
microscopic foundation of the phenomenological Hamiltonian, ${\cal
H}_{\rm WIBM}$, proposed in \cite{Kuwata-Gonokami}.
The agreement of the present theory with experiments supports the
validity of a description of a fermionic system by bosonic fields in
two-dimension, if the excitation is weak. 
This is a strong indication that bosonization can be a powerful tool
also in higher than one dimension.

Helpful discussions with Professor Kuwata-Gonokami and
Dr.\ Suzuura are acknowledged.

\appendix
\section{Different bandstructures: the case of two $s$-type bands}
We have obtained the opposite spin exciton interaction
within the  projection procedure for exciton boson operator including electron and hole spin indices.
The two Hilbert spaces spanned by the excitons corresponding to
left-circularly  and right-circularly polarized photons are completely orthogonal.

One might think that this decoupling is due to the spin
configuration $J^{z}_{e}=\pm1/2$ and
$J^{z}_{h}=\pm3/2$ of the electron and heavy-hole bands
in GaAs.  This is, however, not the case:
We show that for a different band structure with $s$-type
electron and hole bands
($J^{z}_{e}=J^{z}_{h}=\pm1/2$), excitons with $S=\pm$ are also
decoupled before the projection.

First, we define exciton spins in analogy to Eq.\,(\ref{cg-tbl2}).
Since the $J^{z}\equiv J^{z}_{e}+J^{z}_{h}=0$ states are
twofold degenerate in the current case, we consider linear combinations of both
states, which are denoted by $S=0, 0'$.
Now, a spin--matrix is defined by
\begin{equation}
\left(
\begin{array}{c}
|+\rangle \\
|-\rangle \\
|0\rangle \\
|0'\rangle
\end{array}
\right)
=
\left(
\begin{array}{cccc}
1 & 0 & 0 & 0 \\
0 & 1 & 0 & 0 \\
0 & 0 & 1/\sqrt{2} & 1/\sqrt{2} \\
0 & 0 & 1/\sqrt{2} & -1/\sqrt{2}
\end{array}
\right)
\left(
\begin{array}{c}
|+1/2,+1/2\rangle \\
|-1/2,-1/2\rangle \\
|+1/2,-1/2\rangle \\
|-1/2,+1/2\rangle
\end{array}
\right).
\label{cg-tbl1}
\end{equation}
Note that only the spin variables will be different from
the case of $J^{z}_{h}=\pm3/2$, whereas the orbital part remains the same.
The interaction terms of the dipole active $1s$ excitons with $S=\pm$ {\em before} the projection
procedure are obtained as
\begin{equation}
 {\cal H}^{\pm}_{1/2\otimes1/2\,1s}
=\frac{U}{2\Omega}\sum_{S=\pm}\sum_{{\bf k}{\bf k}'{\bf q}}
b^{\dagger}_{{\bf k}+{\bf q}\,S}b^{\dagger}_{{\bf k}'-{\bf q}\,S}b_{{\bf k}'\,S}b_{{\bf k}\,S},
\end{equation}
which correspond to Eq.\,(\ref{Hexpm}).
The interaction strength $U$ is shown in Eq.\,(\ref{1-exchange-orbit}).
Note that the opposite spin exciton interaction is absent as in the case
of $J^{z}_{h}=\pm3/2$.
The interaction term which includes the dipole inactive exciton
corresponding to Eq.\,(\ref{Hex'}) is
\begin{eqnarray}
 {\cal H}^{\prime}_{1/2\otimes1/2\,1s}&=&
\frac{U}{\Omega}\sum_{{\bf k}{\bf k}'{\bf q}}
\biggl[\sum_{S=0,0'}
 \biggl(b^{\dagger}_{{\bf k}+{\bf q}\,+}b^{\dagger}_{{\bf k}'-{\bf q}\,S}b_{{\bf k}'\,S}b_{{\bf k}\,+}
       +b^{\dagger}_{{\bf k}+{\bf q}\,-}b^{\dagger}_{{\bf k}'-{\bf q}\,S}b_{{\bf k}'\,S}b_{{\bf k}\,-}
\biggr)
\nonumber\\
&&+\biggl(
\frac{1}{2} b^{\dagger}_{{\bf k}+{\bf q}\,+}b^{\dagger}_{{\bf k}'-{\bf q}\,-}b_{{\bf k}'\,0}b_{{\bf k}\,0}
-\frac{1}{2}b^{\dagger}_{{\bf k}+{\bf q}\,+}b^{\dagger}_{{\bf k}'-{\bf q}\,-}b_{{\bf k}'\,0'}b_{{\bf k}\,0'}
+{\rm h.c.}
\biggr)
\nonumber\\
&&+\frac{1}{4}b^{\dagger}_{{\bf k}+{\bf q}\,0}b^{\dagger}_{{\bf k}'-{\bf q}\,0}b_{{\bf k}'\,0}b_{{\bf k}\,0}
+\frac{1}{4}b^{\dagger}_{{\bf k}+{\bf q}\,0'}b^{\dagger}_{{\bf k}'-{\bf q}\,0'}b_{{\bf k}'\,0'}b_{{\bf k}\,0'}
+\frac{1}{4}b^{\dagger}_{{\bf k}+{\bf q}\,0}b^{\dagger}_{{\bf k}'-{\bf q}\,0}b_{{\bf k}'\,0'}b_{{\bf k}\,0'}
\nonumber\\
&&+\frac{1}{4}b^{\dagger}_{{\bf k}+{\bf q}\,0'}b^{\dagger}_{{\bf k}'-{\bf q}\,0'}b_{{\bf k}'\,0}b_{{\bf k}\,0}
+b^{\dagger}_{{\bf k}+{\bf q}\,0}b^{\dagger}_{{\bf k}'-{\bf q}\,0'}b_{{\bf k}'\,0'}b_{{\bf k}\,0}
\biggr].
\end{eqnarray}
From a calculation analogous to the one in Sec.\,IV, we obtain the
$1s$ exciton interaction terms after projection.
The projection modifies the dipole active part of the interaction into the form of
\begin{eqnarray}
 \tilde{\cal H}^{\pm}_{1/2\otimes1/2\ 1s}&=&
\frac{U-U'}{2\Omega}\sum_{S=\pm}\sum_{{\bf k}{\bf k}'{\bf q}}
b^{\dagger}_{{\bf k}+{\bf q}\,S}b^{\dagger}_{{\bf k}'-{\bf q}\,S}b_{{\bf k}'\,S}b_{{\bf k}\,S}
\nonumber\\
&&-\frac{U'}{\Omega}\sum_{{\bf k}{\bf k}'{\bf q}}
b^{\dagger}_{{\bf k}+{\bf q}\,+}b^{\dagger}_{{\bf k}'-{\bf q}\,-}b_{{\bf k}'\,-}b_{{\bf k}\,+},
\end{eqnarray}
which corresponds to Eq.\,(\ref{tildeHexpm}).
Here, $U'$ is the contribution from the higher exciton states, which is
the same as Eq.\,(\ref{U'}).
As in the case of $J^{z}_{h}=\pm3/2$, the opposite spin
exciton interaction is present, and the interaction strength
of the equal spin exciton interaction is reduced.
The remaining interaction terms of $1s$ excitons,
corresponding to Eq.\,(\ref{tildeHex'}), are obtained as
\begin{eqnarray}
\tilde{\cal H}^{\prime}_{1/2\otimes1/2\,1s}&=&
\frac{U-U'}{\Omega}\sum_{{\bf k}{\bf k}'{\bf q}}
\sum_{S=0,0'}
 \biggl(b^{\dagger}_{{\bf k}+{\bf q}\,+}b^{\dagger}_{{\bf k}'-{\bf q}\,S}b_{{\bf k}'\,S}b_{{\bf k}\,+}
       +b^{\dagger}_{{\bf k}+{\bf q}\,-}b^{\dagger}_{{\bf k}'-{\bf q}\,S}b_{{\bf k}'\,S}b_{{\bf k}\,-}
\biggr)
\nonumber\\
&&+\frac{U}{2\Omega}\sum_{{\bf k}{\bf k}'{\bf q}}\biggl(
 b^{\dagger}_{{\bf k}+{\bf q}\,+}b^{\dagger}_{{\bf k}'-{\bf q}\,-}b_{{\bf k}'\,0}b_{{\bf k}\,0}
-b^{\dagger}_{{\bf k}+{\bf q}\,+}b^{\dagger}_{{\bf k}'-{\bf q}\,-}b_{{\bf k}'\,0'}b_{{\bf k}\,0'}
+{\rm h.c.}
\biggr)
\nonumber\\
&&+\frac{U-2U'}{4\Omega}\sum_{S=0,0'}\sum_{{\bf k}{\bf k}'{\bf q}}
 b^{\dagger}_{{\bf k}+{\bf q}\,S}b^{\dagger}_{{\bf k}'-{\bf q}\,S}b_{{\bf k}'\,S}b_{{\bf k}\,S}
\nonumber\\
&&+\frac{U}{4\Omega}\sum_{{\bf k}{\bf k}'{\bf q}}
\biggl(
 b^{\dagger}_{{\bf k}+{\bf q}\,0}b^{\dagger}_{{\bf k}'-{\bf q}\,0}b_{{\bf k}'\,0'}b_{{\bf k}\,0'}
+b^{\dagger}_{{\bf k}+{\bf q}\,0'}b^{\dagger}_{{\bf k}'-{\bf q}\,0'}b_{{\bf k}'\,0}b_{{\bf k}\,0}
\biggr)
\nonumber\\
&&+\frac{U-U'}{\Omega}\sum_{{\bf k}{\bf k}'{\bf q}}
b^{\dagger}_{{\bf k}+{\bf q}\,0}b^{\dagger}_{{\bf k}'-{\bf q}\,0'}b_{{\bf k}'\,0'}b_{{\bf k}\,0}.
\end{eqnarray}
\section{estimation of $U'$}
The wave functions of $2p$ states in two-dimension are
\begin{equation}
 \varphi({\bf q})_{2p,\pm}
=\frac{9\sqrt{6\pi}a^{2}_{0}}{\Omega}\frac{q_{x/y}}{[1+(3|{\bf q}|a_{0}/2)^{2}]^{5/2}}.
\end{equation}
Assuming that the transferred momentum in each scattering processes is
small, the ${\bf K}$ summation is replaced by the products of the number of
the states cut-off by $C$ as
$
 \sum_{{\bf K}}f({\bf K})\approx (\Omega C^{2}/a^{2}_{0})\times f(0).
$
The two arguments in the absolute are calculated as
\begin{eqnarray}
&& \sum_{{\bf p},{\bf p}'}\tilde{V}({\bf p}-{\bf p}')
\tilde{\varphi}^{*}_{1s}({\bf p})\tilde{\varphi}^{*}_{1s}({\bf p})
\tilde{\varphi}_{2p}({\bf p})\tilde{\varphi}_{2p}({\bf p}')
=\frac{27}{4096}(1180-819\log 3)\frac{\pi a^{2}_{0}}{2}E^{2D}_{ex},
\\
&& \sum_{{\bf p},{\bf p}'}\tilde{V}({\bf p}-{\bf p}')
\tilde{\varphi}^{*}_{1s}({\bf p})\tilde{\varphi}^{*}_{1s}({\bf p}')
\tilde{\varphi}_{2p}({\bf p})\tilde{\varphi}_{2p}({\bf p}')
=\frac{127575}{32768}\frac{\pi^{3}a^{2}_{0}}{2\xi^{5}}E^{2D}_{ex},
\end{eqnarray}
where
a non-dimensional parameter $\xi$ is introduced for simplification of
the calculation.  
This parameter is determined by the following:  
The integral shown in the left hand side of the next
equation, which is encountered in the above calculation, 
is approximated by the right hand side integral.  
\begin{eqnarray}
\int_{0}^{\infty}dxJ_{1}\left(\frac{r}{a}x\right)\frac{x^2}{(1+x^2)^{3/2}(1+9x^2)^{5/2}}
&\approx&
\int_{0}^{\infty}dxJ_{1}\left(\frac{r}{a}x\right)\frac{x^2}{(1+\xi x^2)^4}
\nonumber\\
&=&\frac{r^{3}}{48\xi^{6}a^{3}}K_{2}\left(\frac{r}{a\xi}\right),
\end{eqnarray}
where $a=a_{0}/2$, and $J_{1}(x)$ and $K_{2}(x)$ are the first order of
Bessel function and the second order modified Bessel functions,
respectively \cite{Gradshteyn}.  
The optimum value of the parameter is $\xi^{2}=5.2$.  
The parameter $\xi$ is introduced just for a convenience of analytic calculation, 
and will not change a qualitative nature of the discussion.  

From these, it is rewritten as
\begin{eqnarray}
 U'\approx \frac{1}{\Omega}\frac{\Omega C^{2}}{a^{2}_{0}}
\times2&\times&\frac{1}{2E^{2D}_{2p}-2E^{2D}_{1s}}
\Biggl|2\sum_{{\bf p},{\bf p}'}\tilde{V}({\bf p}-{\bf p}')
\Bigl[\tilde{\varphi}^{*}_{1s}({\bf p})\tilde{\varphi}^{*}_{1s}({\bf p})
\tilde{\varphi}_{2p}({\bf p})\tilde{\varphi}_{2p}({\bf p}')
\nonumber\\
&&-
\tilde{\varphi}^{*}_{1s}({\bf p})\tilde{\varphi}^{*}_{1s}({\bf p}')
\tilde{\varphi}_{2p}({\bf p})\tilde{\varphi}_{2p}({\bf p}')
\Bigr]
\Biggr|^{2},
\end{eqnarray}
where the first two-times comes from the number of the $2p$ states in
two-dimension.  
Using $E^{2D}_{1s}=-E^{2D}_{ex}$ and $E^{2D}_{2p}=-E^{2D}_{ex}/9$, 
$U'$ is 
estimated as
$U'\approx 16.5a^{2}_{0}C^{2}E^{2D}_{ex}$.

\begin{figure}
\begin{center}
\epsfile{file=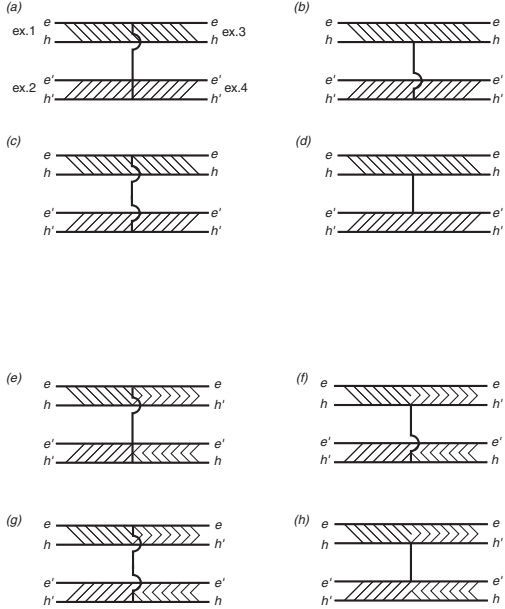,scale=0.4}
\end{center}
\caption{
Diagramatically expression of  direct (a)-(d) and fermionic
exchange interaction (e)-(h) of $1s$ excitons.  The holizontal lines represent an
electron and a hole and the verical line Coulomb interaction of two
particles connected by the line.  
}
\end{figure}
\begin{figure}
\begin{center}
\epsfile{file=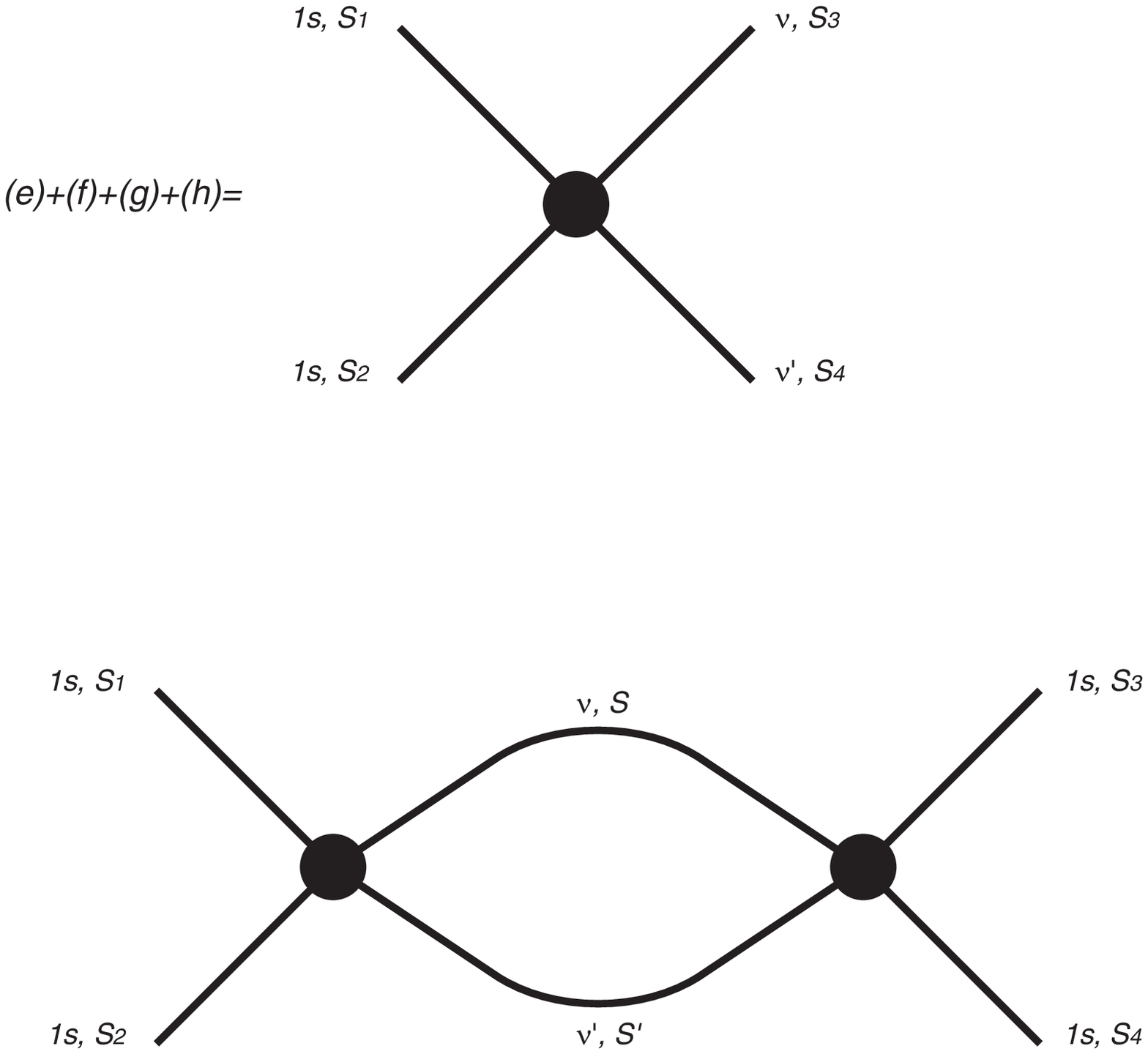,scale=0.4}
\end{center}
\caption
{Successive exchange scattering processes of excitons taken into
the projection procedure.  The bullet shows the sum of the single
exchange scattering processes between the $1s$ and $\nu$ excitons.  
}
\end{figure}

\begin{thebibliography}{99}
\bibitem{short-range}
Actually, our final results are for short-range interaction due to the
	   reason described in the text.
\bibitem{inactive}
However, ${\tilde{\cal H}}^{\prime}_{1s}$, of course, becomes important 
when one measures optical processes in which 
higher-order terms are involved.
\end{thebibliography}
\end{document}